\documentstyle[aps,aipbook,floats,epsf]{revtex}

\begin{document}

\title{A Simultaneous Spectral Invariant Analysis of the GRB
Count Distribution and Time Dilation.}
\author{Ehud Cohen and Tsvi Piran}
\address{Racah Institute of Physics, The Hebrew University, 
Jerusalem, Israel 91904}

\maketitle
\begin{abstract}

The analysis of the BATSE's count distribution within cosmological
models suffers from observational uncertainties due to the variability
of the bursts' spectra: when BATSE observes bursts from different
redshifts at a fixed energy band it detects photons from different
energy bands at the source. This adds a spectral dependence to the
count distribution $N(C)$.  Similarly variation of the duration as a
function of energy \cite{fenimore} at the source complicates the time
dilation analysis.  It has even been suggested that these methods lead
to inconsistent estimates of the redshift from which the bursts are
observed\cite{FenBloom}. Clearly it would be best to combine the
estimates and to perform a joint analysis of the strength and the
duration of the bursts. But for this we have to eliminate first the
spectral dependence problem.

We describe here a new statistical formalism that  performs
the required ``blue shifting" of the count number and the burst
duration in a statistical manner. This formalism allows us to perform
a combined best fit (maximal likelihood) to the count distribution,
$N(C)$, and the duration distribution simultaneously.  The outcome of
this analysis is a single best fit value for the redshift of the
observed bursts.

\end{abstract}

\section*{Introduction}

When BATSE observes bursts from different red-shifts at a fixed energy
band it detects photons from different energy bands at the source.
This spectral dependence complicates the interpretation of the
peak-flux and time-dilation distributions.  So far several attempts
have been made to overcome this problem by modeling the spectral shape
of the bursts. We suggest here a different method which is based on
the availability of multi channel data in different energy bands. The
basic idea beyond our scheme is that we ``view" all bursts at the same
intrinsic energy band independently of their red-shift by by scanning
over the different channels until we find the most likely red-shift
and using this value to ``blue-shift" back the observed spectrum to
the initial spectrum at the source.  Now we look at the same energy
band at the source for all bursts, avoiding the issue of the spectral
shape.

\section*{Standard candle sources.}

Consider first a population of standard candle sources that have a fixed
luminosity, $L$,  at a given energy band centered around an energy $E$. 
For each burst we have a set of measured peak fluxes $C_i(E_i)$ at 
different energy bands.
We determine the
red-shift of a particular burst by solving the equation:
\begin{equation}
L=L(z)=C({E \over 1+z}) d_L^2(z) (1+z)^{-2} \ ,
\end{equation}
where  $d_L$ is the luminosity distance and
$C({E \over 1+z})$ is the observed peak-flux at the energy 
$E \over 1+z$.

One may wonder whether there will be multiple solutions to this
equation. The answer is surprisingly no, provided that the spectrum is
softer than $N(E) dE \propto E^{-1} dE$.  As an example we
consider a simulated source with a spectral shape: $N(E) dE \propto
E^{-\alpha} dE$.  Figure 1 shows the luminosity as calculated by this
method for simulated sources with $\alpha=2.5$ (upper sheet) and
$\alpha=1$ (lower sheet) in an $\Omega=1$ cosmology.  The only $z$
which solves the luminosity equation is the actual redshift.  

\begin{figure}
\begin{center}
\leavevmode
\epsfxsize=300pt \epsfbox{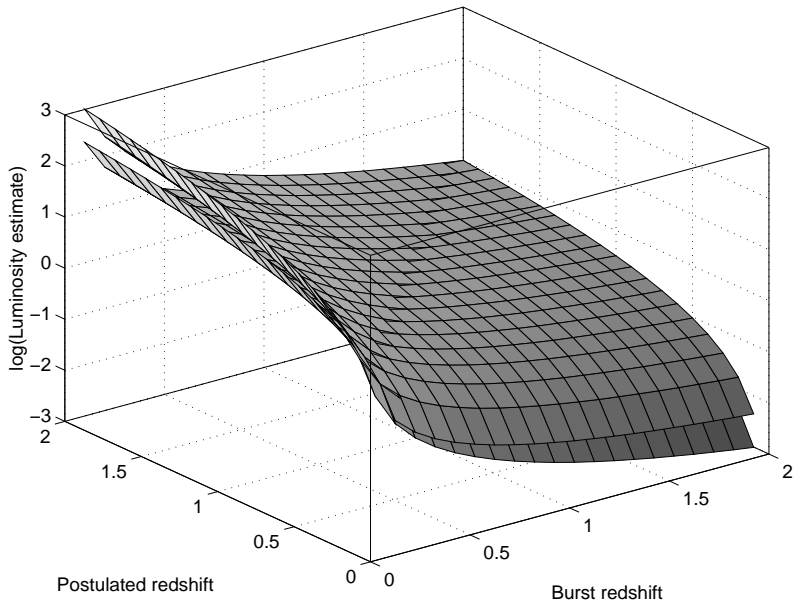}
\end{center}

\caption{The ``blue-shifted" luminosity of an observed burst 
with an intrinsic luminosity $L=1$, with power-law parameter
$\alpha=2.5$ (upper sheet) and $\alpha=1$ (lower sheet) in an
$\Omega=1$ cosmology, for different assumed $z$ values.  The correct
luminosity $L=1$ is obtained only when the chosen $z$ is the same as
the real one.}
\end{figure}

Once we obtain the red-shift for each burst we proceed to compare
the red-shift distribution with the one predicted by a different
cosmological models.  This is done, for example, by  using the
maximum likelihood method.  We check whether the method is
consistent by repeating the whole process  for different energy
channels (at the source).  If there is no noise the different
channels should give the same red-shift distribution.

\section*{Sources with variable luminosity.}

In reality we don't have standard candle sources. Current estimates
suggest that the GRB luminosity function is quite narrow (with
variation of peak flux of no more then one order of magnitude) but it
is unlikely that it is a delta function.  Thus, the luminosity of a
burst is not known a priori, and the redshift can not be deduced
uniquely from the peak flux. However, this method can  be used 
in a statistical manner even when faced with this uncertainty.

We assume that the bursts have a known luminosity function
$\phi(L)$ at a standard energy band at the source.
In fact all that we need to assume is that $\phi(L)$ has
a given functional shape characterized by a few parameters which will
be determined by the analysis.
As in the standard candle case we blue-shift the peak flux 
in each of the energy channels by a suitable factor $(1+z)$
to our canonical energy at the source and we
calculate the corresponding luminosity: $L(z)$, using equation 1. 
We then estimate 
the likelihood that the burst is at redshift $z$i as:
\begin{equation}
h(z)=\phi(L(z)) {dL \over dz} \equiv \tilde\phi(L(z))  \ .
\end{equation}
For  standard candles $\phi(L)=\delta(L-L_0)$ and  
$h(z)=\delta(z-z_0))$.  Figure 2 depicts the spread in $z$ due to a
power-law luminosity distribution:
\begin{equation}
\phi(L)=\left\{ {\matrix{
{\left({L \over L_0}\right)}^3 & L<L_0 \cr
{\left({L \over L_0}\right)}^{-3} & L>L_0 \cr
}} \right .
\end{equation}

\begin{figure}
\begin{center}
\leavevmode
\epsfxsize=250pt \epsfbox{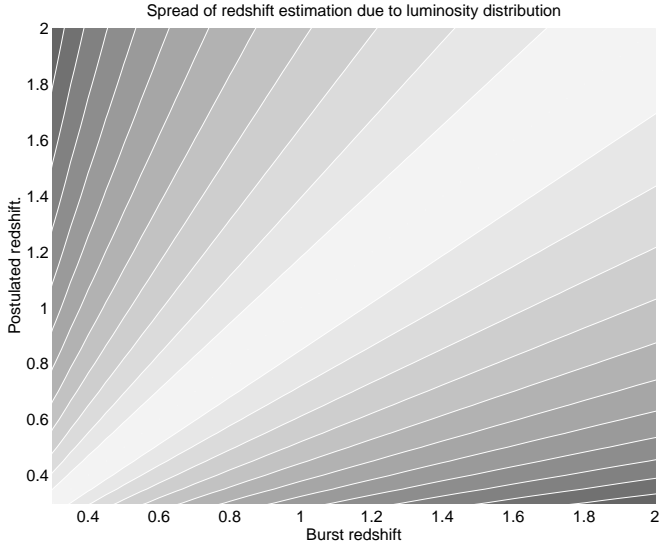}
\end{center}

\caption{Figure 2: Likelihood of a burst from redshift z to be detected as
a nominal burst from a postulated redshift. 
Standard candle sources would make the graph  non zero only
on the main diagonal. 
The spread is due to a power-law luminosity distribution
given by Eq. 3.
The lines represents $33\%$, $10\%$, $3.3\%$... likelihood.
}
\end{figure}

A given cosmological model, with a given evolution model 
(that is number density of bursts per co-moving as a function of cosmological time)
determines $n(z)dz$, the expected number of bursts from a  $dz$ interval
centered around $z$, 
per observer unit time.
With a  given luminosity function
at a fixed energy band at the source we calculate now the likelihood
of a burst, given the set of data $C_i(E_i)$:
\begin{equation}
\matrix{
P(C_{1 \ldots n})= \int  n(z) h(z) dz  \cr
=\sum_i \tilde\phi \left[ {C_i( {E \over 1+z_i} )  d_L(z_i)^2 (1+z_i)^{-2}} \right ] 
n(z_i)  \ . }
\label{like1}
\end{equation}

Once we have this estimate for the likelihood of all bursts we proceed
to write the likelihood function of the whole set of data. Then we 
use the maximal likelihood method to estimate the most likely  
cosmological parameters and the luminosity function parameters.

It is apparent that as the luminosity distribution widens, the
constrain on the redshift weakens, and the information about cosmology
is vaguer.  The difference between reasonable spectrums is not large.
For example to confuse a redshift $z=0.5$ bursts and a redshift $z=1$
with $\alpha=2$ requires a luminosity factor of $4.5$ while $\alpha=1$
requires a luminosity factor of $3.4$. 
Naturally, as the amount of data increases, the confidence 
in the determination of $z$ increases. Once
more, self consistency requires that similar results are obtained when
using different energy channels.

\section*{A joint analysis of time-dilation.}

The  time-dilation analysis should be combined with the peak-flux
estimate.  The combined analysis provide a self consistent method
which estimates the red-shift of each burst using both sets of data.
This
should be more accurate than the standard analysis \cite{norris}
,\cite{band},\cite{wijers} in which the bursts
are grouped into large groups of bright, dim and dimmest bursts.

Using the spectral invariant method we view all bursts in the same
intrinsic energy, and in this way we eliminate the energy dependence
of the bursts' duration.  Now we would like to combine information 
about the duration of a burst in the different energy channels,
$\delta t_i (E_i)$  (this could be  $\Delta t_{50}$ or alternatively
the autocorrelation characteristic time) with the  set 
of peak-fluxes $C_i(E_i)$.
We estimate the likelihood of the burst as:
\begin{equation}
\label{max_like}
\matrix{
P(C_{1 \ldots n},\delta t_{1 \ldots n})= 
\int n(z) h(z) \Psi \left [ {{ \delta t( {E \over 1+z}) \over 1+z },E} \right ] dz = \cr 
\sum_i n(z_i) \tilde\phi \left[ { C( {E \over 1+z_i} )  d_L(z_i)^2 (1+z_i)^{-2} ) } \right ]
\Psi \left [{{\delta t_i({E \over 1+z_i}) \over 1+z_i}, E} \right ] }
\end{equation}
Where $\Psi(t,E)$ is the intrinsic energy dependent
duration distribution function. 
Using the maximum likelihood method we can now estimate the best-fit
parameters of the luminosity function, $\tilde \phi$, the duration
distribution, $\Psi(t,E)$, and the maximal redshift from which the
bursts are observed.

\section*{Discussion}

The spectrum invariant
method is the only way to have a complete usage of all the information
available in the multi-spectral channel
detectors in situations in which the spectra is being red-shifted 
differently from
burst to burst.  It also uses the time-dilation and the
peak-luminosity  in a way that more than just brings a consistent
answer, it uses the whole data to enlarge the statistical confidence
of the results. A luminosity distribution  influences considerably
the confidence the red-shift determination of a single burst origin.
However, it has a minor influence when using a large  number of bursts.
Similarly the duration distribution makes it impossible to determine
the red-shift of a single burst, however one can determine the 
parameters of the duration distribution and find whether the red-shift data
determined from the time dilation analysis is consistent with the
one determined from the peak-flux distribution.

\end{document}